\documentclass{article}
\usepackage[utf8]{inputenc}
\usepackage[english]{babel}

\usepackage[letterpaper,top=2cm,bottom=2cm,left=3cm,right=3cm,marginparwidth=1.75cm]{geometry}

\usepackage{amsmath,amssymb}
\usepackage{braket}
\usepackage{graphicx}
\usepackage{algorithm}
\usepackage[colorlinks=true, allcolors=blue]{hyperref}
\usepackage{authblk}
\usepackage[toc,page]{appendix}
\usepackage{listings}
\usepackage{xcolor}
\usepackage{tabularx}
\usepackage{multicol}
\usepackage[htt]{hyphenat}
\usepackage[hyphenbreaks]{breakurl}

\definecolor{codecomment}{RGB}{64, 128, 128}
\definecolor{codegreen}{RGB}{0, 128, 0}
\definecolor{backcolour}{RGB}{247, 247, 247}
\definecolor{codered}{RGB}{186, 33, 33}
\lstdefinestyle{python}{
    backgroundcolor=\color{backcolour},   
    commentstyle=\color{codecomment},
    keywordstyle=\color{codegreen},
    stringstyle=\color{codered},
    basicstyle=\ttfamily\footnotesize,
}

\DeclareUnicodeCharacter{211D}{$\mathbb{R}$}

\title{\texttt{trainsum} --- A Python package for quantics tensor trains}
\author[a,b]{Paul Haubenwallner\footnote{\href{mailto:paul.haubenwallner@igd.fraunhofer.de}{paul.haubenwallner@igd.fraunhofer.de}}}
\author[a,b]{Matthias Heller\footnote{\href{mailto:matthias.heller@igd.fraunhofer.de}{matthias.heller@igd.fraunhofer.de}}}
\affil[a]{Fraunhofer Institute for Computer Graphics Research IGD, Darmstadt, Germany}
\affil[b]{Technical University of Darmstadt, Interactive Graphics Systems Group, Darmstadt, Germany
}
\date{}

\begin{document}

\maketitle
\vspace{-1cm}

\begin{abstract}
We present \texttt{trainsum}, a versatile Python package for doing computations with multidimensional quantics tensor trains: \url{https://github.com/fh-igd-iet/trainsum}.
Using the Array API standard~\cite{ArrayStandrd} together with \texttt{opt\_einsum}~\cite{G.A.Smith2018}, \texttt{trainsum} allows the effortless approximation of tensors or functions by tensor trains independent of their shape or dimensionality.
Once approximated, our package can perform normal arithmetic operations with quantics tensor trains, including addition, Einstein summations and element-wise transformations.
It can be therefore used for generic computations with applications in simulation, data compression, machine learning and data analysis.
\end{abstract}

\tableofcontents

\section{Introduction}
\label{sec:intro}
Tensor networks have become an important tool in quantum physics, mathematics and computer science  \cite{Banuls2023, Wang2023, Berezutskii2025} for their ability to approximate large tensors by using smaller tensors, also called cores, together with a prescribed contraction scheme.
In other words, tensor networks approximate ``standard'' tensors by adding internal structure.
This structure is not unique and a plethora of different types are well studied in literature \cite{Schollwoeck2011a, Vidal2007, Dolgov2013, Kressner2014}.
The networks can be roughly divided into two groups: Tensor networks that allow for a normalized or canonical form and those, that do not.
The canonical form is a well defined state, where the norm of a tensor network only depends on a single core, which is the prerequisite for certain algorithms like the density matrix renormalization group (DMRG) \cite{White1992} or the tensor cross interpolation (TCI) \cite{OSELEDETS201070} to be applicable.
All tensor networks that do not have cycles in their contraction path possess canonical formats.

Among the most studied structures are linear tensor networks which are called tensor trains in the math community or matrix product states in the physics community \cite{Schollwoeck2011a, Oseledets2011}.
Initially used to simulate spin-chains \cite{White1992}, they attract an increasing amount of interest due to the fast evolving field of quantum computing (QC).
Tensor trains are well suited to approximate quantum states and are therefore widely applied for the simulation of quantum circuits \cite{Banuls2023}.
The entanglement of a quantum state is directly linked to the size of the cores of the tensor train, such that it is possible to evaluate if a circuit is easy to simulate classically or not \cite{Berezutskii2025}.
Another similarity between tensor trains and qubits is the inability to access all properties of the underlying function efficiently.
For example, the access to the full state, which in certain cases is highly desirable, is not possible without a significant amount of measurements in the QC context or a significant amount of computation for tensor networks.
In contrast to that, some properties like the computation of an expectation value of some operator can be done with very little effort.

The reason for tensor trains to be used a lot in quantum physics lies within the fact that highly dimensional tensors naturally arise there,
while other fields quite often deal with problems with a small number of dimensions which however have large sizes.
With the quantics formalism \cite{Khoromskij2011} it is possible to utilize the strengths of tensor networks also for these problems.
In essence, the quantics formalism factorizes each dimension into multiple smaller ones allowing in many cases an effective approximation as tensor network.
As most software packages tend to prioritize the simulation of quantum states \cite{Fishman2022, gray2018quimb}, there is no easy-to-use, open-source package for doing arithmetic and solving discretized problems with quantics tensor trains.
Notable exceptions that cover some aspects are \cite{TT_box, NunezFernandez2025, tntorch, torchtt}.

With this in mind we developed \texttt{trainsum}, a Python package allowing to perform calculations with $N$-dimensional quantics tensor trains.

\paragraph{Novelty of our approach}
As the name suggests, \texttt{trainsum}'s main feature is the capability to write linear transformations of quantics tensor trains in the Einstein notation similar to, e.g., NumPy's einsum function.
Since the ranks of a tensor network rise exponentially with the number of such linear transformations performed, \texttt{trainsum} is not only capable to execute such operations exactly but also approximately.
There are two main algorithms for reducing the ranks of an operation.
The first one is the so called zip-up algorithm, where local approximate matrix decompositions lead to an approximation.
The second one comprises DMRG-like variational algorithms that try to reduce the distance between the approximation and the exact result.

In addition to that, \texttt{trainsum} allows the effortless definition of many important structured tensors as quantics tensor trains including some trigonometric functions, polynomials, shift matrices or discrete fourier transformations.
Since not all functions have a simple analytical definition the tensor cross interpolation algorithm is also implemented for converting any kind of data or function that can be sampled into a tensor train.

One feature that also stands out from other packages is the factorization of a dimension.
In the literature many algorithms and concepts are only defined for dimensions that have a size $2^N$ and can therefore be factorized into $N$ dimensions of size two.
\texttt{trainsum}'s approach is more flexible, in that any factorization is allowed as long as the original size is reproduced.
This makes it well suited for problems that include tensors with arbitrary dimensions where the size is not a power of 2 as is often the case.

The remainder of this article is structured as follows.
In Section \ref{sec:notation} we describe our basic notation, followed by Section \ref{sec:arithmetic}, where we show how tensor train arithmetic works.
Subsequently we provide in Section \ref{sec:structured_tensors}  some explicit formulas for important functions and operations directly as tensor trains together with a short paragraph about solvers and other algorithms in Section \ref{sec:solver}.
After that, we provide a user guide for \texttt{trainsum} in Section \ref{sec:user_guide} and a conclusion in Section \ref{sec:conclusion}.

\section{Notation} \label{sec:notation}
In  the following, we are dealing with tensors with external indices, which describe high-dimensional data.
We write $A(i)$ to denote a tensor with one dimension, i.e., a tensor with one index $i$ with $0< i\leq{\rm dim}(i)$.
In order to approximate such tensor by factorizing it into smaller constituents, we first need to factorize the index $i$ using integer fractions:
\begin{equation}
    A(i) \equiv A(i_1,\dotsc i_n),
\end{equation}
where $i$ depends on $i_q$
\begin{equation}
i=\sum^n_{q=1} \left(\prod_{r={q+1}}^n b^i_r\right)\cdot i_q\equiv \sum^n_{q=1} c^i_q i_q,
\label{eq:quantization}
\end{equation}
where we defined the coefficients $c^i_q=\prod_{r={q+1}}^n b^i_r$, which we will use frequently in the following.
The integer bases $b^i_q$ are chose such that
\begin{equation}    
{\rm dim}(i) = c_0 = \prod^n_{q=1} b^i_q.
\end{equation}
This operation corresponds to reshaping an array into another one with more, but lower-dimensional indices, which we will call factorizing a dimension or an index.

Note that compared to many other existing packages, we allow flexibility regarding the dimension of the data, that the tensor trains describes.
For instance, given a tensor of shape $(i,j,\dotsc)$, one can rewrite this as a tensor train with dimensions $((i_1,i_2,\cdots i_{n_1}),(j_1,\dotsc j_{n_2})\dotsc)$, where $i=\prod_{q=1}^{n_1}c^i_q i_{q}$ and $j=\prod_{q=1}^{n_2}c^j_q j_{q}$.
This introduces some difficulties in the algorithms and operations that have to be implemented for tensor trains that will be explained in what follows.
While a generalization of some algorithms like tensor cross interpolation (TCI) or density matrix renormalization group (DMRG) are straightforward, operators that were explicitly constructed for the case where the dimension is a power of two have to be adopted.

Having factorized $A(i)$ one can then use low-rank factorization on the tensor $A(i_1,\dotsc i_n)$ to get the tensor train and we will write
\begin{equation}
    A(i)= A(i_1,\dotsc i_n) = C_1(i_1) \cdot C_2(i_2) \cdots C_n(i_n),\label{eq:vector_dec}
\end{equation}
where $C_q(i_q)$ are the cores (depending on the indices $i_q$) and define the tensor train decomposition of $A$.
For each index $i_q$, these cores are $d_q\times d_{q+1}$ matrices, where $d_q$ and $d_{q+1}$ are the bond dimensions or ranks of the tensor train.
Note that the above notation can also be extended to tensors with more indices. 
For instance, a tensor with three dimensions, i.e., $A(i,j,k)$ can be decomposed as
\begin{equation}
    A(i,j,k) = C_1(i_1,j_1,k_1) \cdot C_2(i_2,j_2,k_2) \cdots C_n(i_n,j_n,k_n).\label{eq:tensor_dec}
\end{equation}
While Eqs.~\eqref{eq:vector_dec} and  \eqref{eq:tensor_dec} are understood for fixed indices (i.e., fixing the specific index also fixes the cores in the decomposition) it is sometimes necessary to write down the full cores for all external indices at once.
This can be done by utilizing the rank product.
For this, we first rewrite the equations by introducing explicitly indices for the bond dimensions:
\begin{equation}
    A(i,j,\dotsc) =\sum_{\mu_1,\dotsc,\mu_n} C^{\mu_1}_1(i_1,j_1,\dotsc) \; C^{\mu_1\mu_2}_2(i_2,j_2,\dotsc)\; \dotsc C^{\mu_{n-1}}_n(i_n,j_n,\dotsc),
\end{equation}
which is again valid, for any fixed index $(i,j,\dotsc)$. 
To define the the full tensor train array at once, i.e., for all external and internal indices, we have to introduce the rank product \cite{Kazeev2013},
\begin{align}
    A = 
    &\left(\begin{array}{c}
        C_1^{1}(i_1,j_1,\dotsc) \\
        C_1^{2}(i_1,j_1,\dotsc)\\
        \vdots \\
        C_1^{d_1}(i_1,j_1,\dotsc)\\
    \end{array}\right) \boxtimes  \left(\begin{array}{ccc}
        C^{11}_2(i_2,\dotsc) & \dotsc & C^{1d_2}_2(i_2,j_2\dotsc) \\
        \vdots & \cdots & \vdots \\
        C^{d_11}_2(i_2,j_2\dotsc) & \dotsc & C^{d_1d_2}_2(i_2,j_2\dotsc) \\
\end{array}\right)\boxtimes\cdots \left(\begin{array}{c}
        C_n^{1}(i_n,j_n\dotsc) \\
        C_n^{2}(i_n,j_n\dotsc)\\
        \vdots \\
        C_n^{d_1}(i_n,j_n\dotsc)\\
    \end{array}\right)^T,\label{eq:rank_prod}
\end{align}
where the rank product between cores, denoted by $\boxtimes$, is defined as
\begin{align}
&    C_{q}\boxtimes C_{q+1} = \left(\begin{array}{ccc}
        C_q^{11} & \dotsc & C_q^{1d_2} \\
        \vdots & \cdots & \vdots \\
        C_q^{d_11} & \dotsc & C_q^{d_1d_1} \\
    \end{array}\right)\boxtimes
    \left(\begin{array}{ccc}
        C_{q+1}^{11} & \dotsc & C_{q+1}^{1d_3} \\
        \vdots & \cdots & \vdots \\
        C_{q+1}^{d_21} & \dotsc & C_{q+1}^{d_2d_3} \\
    \end{array}\right)=\nonumber\\[0.5em]
    &\left(\begin{array}{ccc}
        C_q^{11}\otimes C_{q+1}^{11}+C_q^{12}\otimes C_{q+1}^{21} +\cdots +C_q^{1d_2}\otimes C_{q+1}^{d_21} & \cdots & C_q^{11}\otimes C_{q+1}^{1d_3} +\cdots +C_q^{1d_2}\otimes C_{q+1}^{n_2n_3} \\
        \vdots & \cdots & \vdots \\
        C_q^{d_11}\otimes C_{q+1}^{11}+C_q^{d_12}\otimes C_{q+1}^{21} +\cdots +C_q^{d_1d_2}\otimes C_{q+1}^{d_21} & \dotsc & C_q^{d_11}\otimes C_{q+1}^{1d_3} +\cdots +C_q^{d_1d_2}\otimes C_{q+1}^{d_2d_3} \\
    \end{array}\right),
\end{align}
where all entries $C_q^{\mu_q \mu_{q+1}}$ and $C_{q+1}^{\mu_{q+1} \mu_{q+2}}$ are multidimensional arrays (as they also depend on some external indices) and $\otimes$ is the usual outer product on tensors.

%Tensors with higher dimensionality define linear maps, which can be defined in terms of %Einstein summations.
%For instance, one can define a matrix-vector product as follows
%\begin{equation}
%    y(i) = \sum_{j} A(i,j)\; x(j),
%\end{equation}
%where $A(i,j)$ is tensor with two dimensions (matrix) and $y(i)$ and $x(j)$ are tensors with %one dimensions each (vectors).
%In Python most libraries dealing with arrays (like NumPy or Torch) allow the usage of Einstein %summation.
%The above equation would be implemented as
%\begin{lstlisting}[language=Python]
%import numpy as np
%A = np.rand(10, 10)
%x = np.rand(10)
%y = np.einsum('ij,j->i', A, x)
%\end{lstlisting}
%\texttt{trainsum} allows the user to define such maps via Einstein summation using the same syntax directly on the level of the tensor trains.

Since \texttt{trainsum} is mainly designed for the low-rank approximation of multi-variate functions defined on a grid, we have to introduce additional notation.
%Given a domain of real numbers $x\in [a,b]$ with $b>a$ and some function $f(x)$, we discretize the domain into a uniform grid with $N$ points.
%We can factorize $N$ into $n$ integers (e.g, by using a prime decomposition), such that $N=\prod_{q=1}^n b_q$.
%This leads to the following discretization of $x\in [a,b]$:
A domain of real numbers given by $x\in [a,b]$ with $b>a$, can be discretized into a uniform grid with $N$ points.
To describe the positions of each point with the vector $x(i)$, we write
\begin{equation}
    x(i) = \frac{b-a}{\text{dim}(i)-1}\ i + a.
\end{equation}
Reshaping index $i$ leads to

\begin{align}
    x(i_1,...i_n) &= \frac{b-a}{c^i_0-1}\ \sum^n_{q=1} c^i_q i_q + a.
\end{align}
Introducing $x_q=\frac{b-a}{c^i_0-1} i_q$ and $c^x_q=c^i_q$ we arrive at
\begin{equation}
    x(x_1,...x_n) = \sum^n_{q=1} c^x_q x_q + a.
\end{equation}
We use the notation $x(x_1,\dotsc x_n)$ to indicate that $x$ is now represented as a quantized tensor with values $x_q=\frac{b-a}{c_0-1}\times[0,1,\dotsc b^i_q-1]$.
A generalization of this to multi-variate functions is straightforward. 
For instance, with $f(x,y)$ we introduce two sets of indices.

\section{Arithmetic with tensor trains} \label{sec:arithmetic}
The arithmetic of tensor networks is not at all straightforward and uses multiple algorithms and methods. 
Some methods only work for linear operations while others also work for non-linear ones.
Since all contractions defining a tensor train are linear, there exists an exact solution for all linear operations.
The problem with the exact solution is, that the ranks are rapidly increasing with every operation performed.
A few of such operations can lead to ranks, where the tensor train approximation is less favorable than working with the full tensor.
To counteract the increasing ranks the operations can be approximated using either decomposition (also called zip-up) algorithms or variational algorithms. 
For non-linear operations, it is not possible to write down the exact solution, so that only approximate algorithms are applicable.
Here the so called cross interpolation can be employed.
Table \ref{tab:arithmetic} shows which operation can be performed with which algorithm.

\begin{table}[b!]
    \centering
    \newcolumntype{C}{>{\centering\arraybackslash}X}
    \begin{tabularx}{\textwidth}{c|C|C|C|C}
       operation                & exact         & decomposition & variational   & cross         \\ \hline
       einsum                   & \checkmark    & \checkmark    & \checkmark    & \checkmark    \\
       addition                 & \checkmark    & \checkmark    & \checkmark    & \checkmark    \\
       element-wise functions   &               &               &               & \checkmark    \\
    \end{tabularx}
    \caption{Assignment of the operation types to the different methods.}
    \label{tab:arithmetic}
\end{table}

\subsection{Exact operations}
Every linear operation between two tensor trains has an exact definition.
We distinguish between additions and Einstein summations (which include multiplications or matrix-vector contractions).
An element-wise addition requires both operands to have the same set of dimensions with the same layout describing how the factorized integers are distributed over the tensor cores.
The result of an addition is a tensor train, where the ranks of each core are the sum of the ranks of the corresponding operand cores.
The result cores have a block-like structure along the bond dimensions, where each block corresponds to one of the operand cores \cite{Schollwoeck2011a}.
This structure prevents the interaction between the operand blocks and therefore resembles the exact summation, which can be written as

\begin{align}
    A(i_1)...A(i_n) + B(i_1)...B(i_n)  &= \left(\begin{array}{cc}
        A(i_1) & 
        B(i_1)
        \end{array} \right) \left(
        \begin{array}{cc}
        A(i_2) & 0   \\
        0      & B(i_2)
        \end{array} \right) \cdots \left(
        \begin{array}{cc}
        A(i_n)   \\
        B(i_n)
        \end{array} \right).
\end{align}

\begin{figure}[b!]
    \centering
    \includegraphics[width=0.9\linewidth]{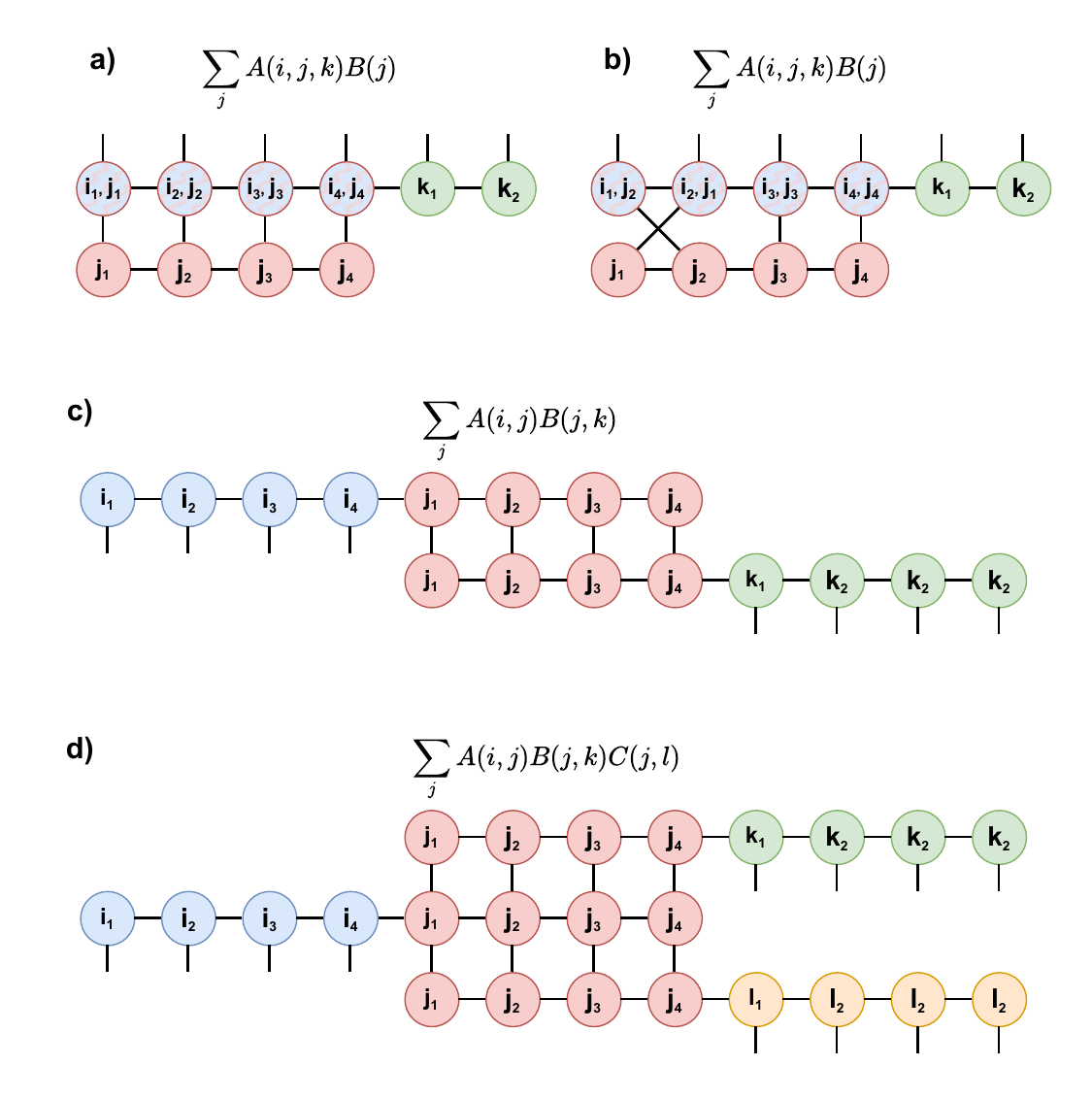}
    \caption{Examples for tensor train shapes for conforming and non-conforming Einstein summations. a) and c) are valid since the the result is a linear tensor network and the $j_q$ dimensions fully overlap. b) is not valid due to a missing overlap of $j_q$. In d) all $j_q$ are overlapping, but the result is not linear.}
    \label{fig:einsum_shapes}
\end{figure}

Turning towards Einstein summations, it is very important to stress that there is an unambiguous mapping of such a summation using normal dimensions to a summation with factorized dimensions.
If a dimension is contracted so are all components of its factorized counterpart.
When writing for example a matrix-vector multiplication
\begin{align}
    C(i) = \sum_j A(i,j) B(j)
\end{align}
a factorization of $i$ into $\prod^{n_1}_{q=1} c^i_qi_q$ and $j$ into $\prod^{n_2}_{q=1} c^j_qj_q$ leads to the summation
\begin{align}
    C(i_1...i_{n_1}) = \sum_{j_1...j_{n_2}} A(i_1...i_{n_1}j_1...j_{n_2}) B(j_1...j_{n_2}).
    \label{eq:quant_sum}
\end{align}
The shape requirements of the operands are different with respect to an addition, since the operands do not even have to share the same dimensionality.
They are depicted graphically with the Penrose notation in Figure \ref{fig:einsum_shapes}.
An Einstein summation requires partly matching shapes in those regions where factored indices are contracted or multiplied.
Additionally the contractions must lead to a result which again represents a linear tensor network.
While these requirements are not strictly needed to perform the overall Einstein summation, which in the worst case would require to compute the full tensors, they lead to efficient contraction schemes and therefore allow a fast computation.

Eq.~\eqref{eq:quant_sum} together with the shape requirements allow the description of the full Einstein summation as a sequence of smaller Einstein summations.
Since the inner bond dimensions are not part of the contractions, they are not affected and appear in the local results.
With the guarantee that the result itself is again a tensor train, the bond dimensions of the result can be grouped together to one left bond and one right bond dimension.
Mathematically the summation of one factorized dimension $i$ can be written as
\begin{align}
    \sum_{i_1,...i_n} A(i_1,...)...A(i_n,...) B(i_1,...)...B(i_n,...) 
        %=& \sum_{i_1...,i_n} A(i_1,...,i_{n_1},...) B(i_1,...,i_{n1},...) \nonumber \\
        =& \sum_{i_1...,i_n} A^{\mu_1}(i_1,...) B^{\nu_1}(i_1,...) \nonumber\\
         & \sum_{i_1...,i_n} A^{\mu_1\mu_2}(i_2,...) B^{\nu_1\nu_2}(i_2,...)\ ...\nonumber\\
         & \sum_{i_1...,i_n} A^{\mu_{n-1}}(i_n,...) B^{\nu_{n-1}}(i_n,...)\nonumber\\
        =& C^{\xi_1}(i_1,...) 
           C^{\xi_1\xi_2}(i_2,...)\ ...
           C^{\xi_{n-1}}(i_n,...)
    \label{eq:einsum}
\end{align}
with $C^{\xi_{q},\xi_{q+1}}(i_q,...)$ as the result of the local summation $\sum_{i_q} A^{\mu_{q},\mu_{q+1}}(i_q,...) B^{\nu_{q},\nu_{q+1}}(i_q,...)$ together with a flatten operation of the left and right bond dimensions so that $\xi_q=\overline{\mu_q\nu_q}$.
If the Einstein summation contracts multiple dimensions at once or defines multiplications the only thing that needs to be adjusted is the corresponding summation.

\subsection{Decomposition algorithms}
\begin{figure}[t!]
    \centering
    \includegraphics[width=0.9\linewidth]{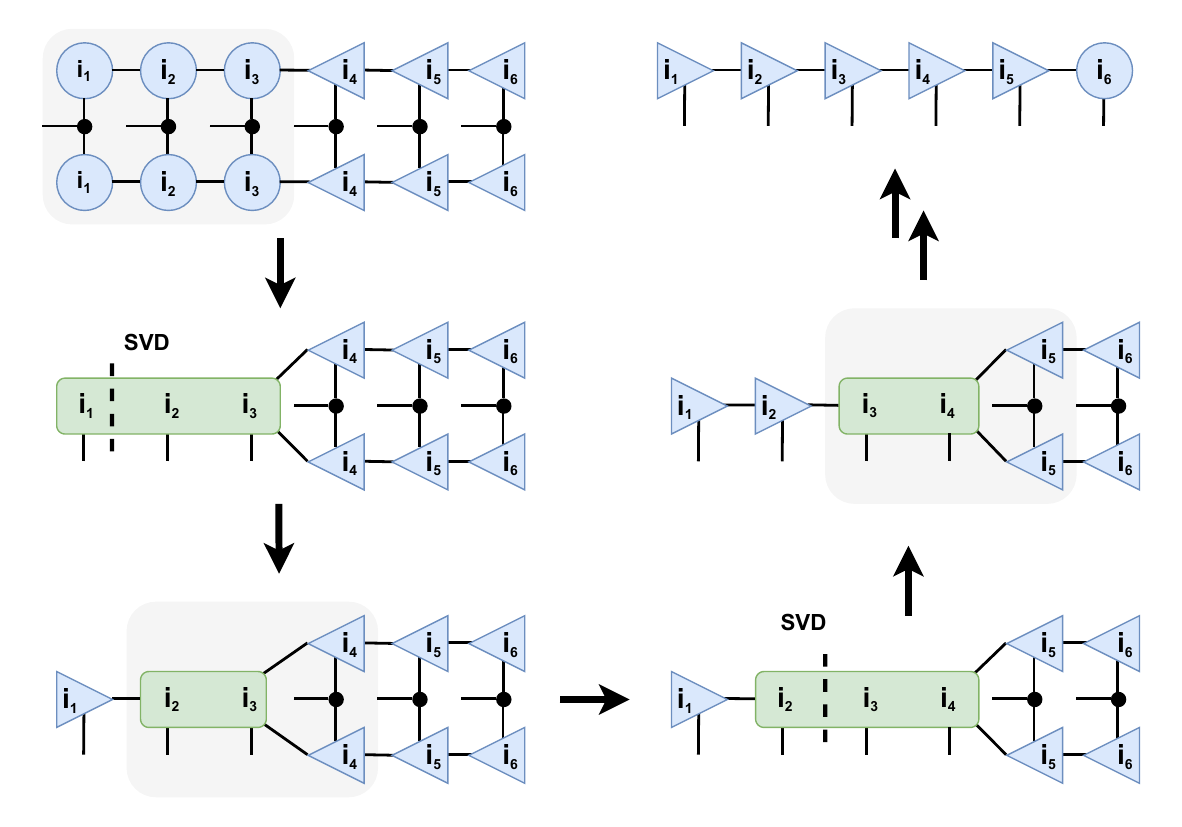}
    \caption{Decomposition algorithm for a multiplication of two tensor trains using three cores at each step. The black dots are multidimensional Kronecker deltas. The gray background indicates a contraction over all closed dimensions, while the dotted lines stand for a singular value decomposition. Orthonormal cores are depicted as triangles where the direction of the orthonormalization is the tip of the triangle.}
    \label{fig:decomposition_algorithm}
\end{figure}

Decomposition-like algorithms, often also called zip-up algorithms \cite{Stoudenmire2010, Paeckel2019}, use approximate matrix decompositions to truncate the ranks of an exact operation.
The exact operation that is actually performed is irrelevant.
The relevant point is that an exact operation can be defined.

Before the algorithm starts, the tensor trains that are part of the operation shift their normalization center to the left end.
Then, the first $N$-cores ($1\leq N\leq n$) are contracted exactly according to some calculation rule and along the resulting bond dimensions.
For example, considering an operation acting on two one-dimensional tensor trains, $A(i_1)A(i_2)...(A(i_n)$ and $B(i_1)B(i_2)...B(i_n)$, with the result $C(i_1)C(i_2)...C(i_n)$, the newly created super-core would be $C(i_1,...,i_N)$.
Since the normalization centers were shifted to the left, the norm of the result only depends on $C(i_1,...,i_N)$.
An approximation for the first core can be done by splitting $C(i_1,...,i_N)$ into $C(i_1)C(i_2,...,i_N)$ with some kind of approximate matrix decomposition.
Depending on the decomposition, the normalization onto $C(i_2,....,i_N)$ might be lost.
If that happens, it needs to be restored, for example using a QR decomposition of $C(i_1)$ and a contraction of $R$ with $C(i_2,....,i_N)$.

$C(i_2,....,i_N)$ can then be extended to $C(i_2,....,i_{N+1})$ using $A(i_{N+1})$ and $B(i_{N+1})$ with a subsequent decomposition into $C(i_2)$ and $C(i_3,...,i_{N+1})$ followed by a normalization onto $C(i_3,...,i_{N+1})$.
This process is repeated until no more cores can be added, after which the super-core is fully decomposed.
The algorithm is shown graphically in Figure \ref{fig:decomposition_algorithm}.

\subsection{Variational algorithms}
Variational algorithms try to minimize the distance between the exact result of some operation $\hat{O}$ and an initially guessed state \cite{Schollwoeck2011a, Paeckel2019}.
For some operation $\hat{O}$ acting on two one-dimensional tensor trains $A(i_1)A(i_2)...(A(i_n)$ and $B(i_1)B(i_2)...B(i_n)$, with the initial guess $C(i_1)C(i_2)...C(i_n)$, the minimization can be expressed as
\begin{equation}
    \left| C - \hat{O}(A, B) \right| = 0,
\end{equation}
which can be rewritten as
\begin{equation}
    \braket{C|C} + \braket{\hat{O}(A, B)|\hat{O}(A, B)} - \braket{C|\hat{O}(A, B)} - \braket{\hat{O}(A, B)|C} = 0.
\end{equation}
Deriving with respect to $C^*(i_q)$ of $\bra{C}$, the equation turns into
\begin{equation}
    \frac{\partial}{\partial C^*(i_q)} \braket{C|C} = \frac{\partial}{\partial C^*(i_q)} \braket{C|\hat{O}(A, B)},
\end{equation}
since $C(i_q)$ only appears in $\bra{C}$.
If $C(i_q)$ is the normalization center, it is defined by
\begin{equation}
    C(i_q) = \frac{\partial}{\partial C^*(i_q)} \braket{C|\hat{O}(A, B)}.
    \label{eq:variational}
\end{equation}
Using this construction, $C$ can be optimized variationally core by core.
Usually the start is to the left and the cores are optimized sequentially by going to the right and back again.
A good starting guess for $C$ is the result of the decomposition approach.
By fusing multiple adjacent $C(i_q)$ into a super-core $C(i_q,...,i_{q+N})$ and solving \eqref{eq:variational} with it, it is possible to realize multi-core strategies, which normally leads to better convergence and an efficient way of controlling the ranks of the cores.
After the optimization $C(i_q,...,i_{q+N})$ can be again fully decomposed into $C(i_q)C(i_{q+1})...C(i_{q+N})$.

\subsection{Cross interpolation}
For an introduction to cross interpolation algorithms we refer to \cite{NunezFernandez2025}.
In principle, it is an extension of the CUR matrix decomposition onto tensor trains.
Instead of columns and rows, the algorithm needs access to one-dimensional slices of a multidimensional tensor.
This turns it into some kind of sampling algorithm where the only prerequisite for it to work is a multidimensional function that can be evaluated efficiently.

The main purpose of the cross interpolation is the construction of tensor trains using some generic function, like $\frac{1}{|r|}$, without the need to do a lot of math as shown in section \ref{sec:structured_tensors}.
Besides that, it can be also used for operations, like additions or multiplications.
The reason lies within the fact that tensor trains allow for an efficient sampling of their entries.
Because of that, the cross interpolation is most convenient for element-wise operations performed on the entries on a tensor train.
The element-wise operation can be simply applied to the sampled values and therefore make the cross interpolation see a function of the desired properties.

It is also possible to exactly sample Einstein summations by fixing the result indices of \eqref{eq:einsum}.
This makes the cross interpolation a rather universal algorithm to perform tensor train arithmetic.

\section{Structured tensors as tensor trains} \label{sec:structured_tensors}
Having introduced the arithmetic, we show in the following subsections how to generalize the explicit construction of some important functions as tensor trains, that were previously developed for dimensions which are powers of two. We show exact decompositions for some basic functions ($\sin(x)$, $\cos(x)$ and polynomials), shift matrices, Toeplitz tensors and for the discrete Fourier transformation.

\subsection{Construction of basic functions}
\paragraph{Exponential functions}
We start by deriving the tensorized structure of an exponential function, which, for dimensions that are powers of two, has been considered in \cite{Khoromskij2011}.
Let $(x_1\dotsc x_n)$ be the decomposition of $x\in [a,b]$, as described above.
Then
\begin{align}
    e^{v(x-x_0)} &\approx e^{v \sum^n_{q=1} c^x_q x_q+va-vx_0} \nonumber\\
                        &= e^{v(a-x_0)}\; \prod^n_{q=1} e^{v c^x_q x_q}
\end{align}
where $\approx$ highlights the use of the discretization.
Since the function completely factorizes into a product, in which every term only depends on a single $x_q$, the rank of the resulting tensor train is one.
Concretly, the tensorized structure of $\exp\left(vx\right)$ takes the form
\begin{equation}
    e^{v(x-x_0)}   = \left(\begin{array}{c}
        e^{vc^x_1x_1+va-vx_0}
    \end{array}\right)\cdot
    \left(\begin{array}{c}
        e^{vc^x_2x_2}
    \end{array}\right)\cdots \left(\begin{array}{c}
        e^{vc^x_nx_n}
    \end{array}\right).\label{eq:exp_dec}
\end{equation}
\paragraph{Trigonometric functions}
We continue with the tensor train structure for $\cos(x)$.
Using the same $x$ as before
\begin{align}
    \cos\left(v (x-x_0)\right) 
    &= \frac{1}{2}(e^{iv (x-x_0)}+e^{-i v (x-x_0)}) \nonumber\\
    &\approx \frac{1}{2}\left[\prod^n_{q=1} e^{i\tilde{x}_q} + \prod^n_{q=1} e^{-i\tilde{x}_q}\right]
\end{align}
with $\tilde{x}_q = vc^x_qx_q+\frac{v}{n}(a-x_0)$.
Both terms in the bracket correspond to rank-1 tensor trains. 
Performing the exact addition of these trains, we can directly read-off the decomposition
\begin{align}
    \cos\left(v (x-x_0)\right) &\approx
    \frac{1}{2}\left[ \left(\begin{array}{cc}
        e^{i\tilde{x}_1} & 
        e^{-i\tilde{x}_1}  
    \end{array}\right)\cdot
    \left(\begin{array}{cc}
        e^{i\tilde{x}_2} & 0 \\
        0 & e^{-i\tilde{x}_2} 
    \end{array}\right)\cdots \left(\begin{array}{cc}
        e^{i\tilde{x}_n} \\
        e^{-i\tilde{x}_n}
    \end{array}\right)\right]\nonumber\\
    &=\frac{1}{2}\left[ \left(\begin{array}{cc}
        1& 
        1
    \end{array}\right)\cdot
    \left(\begin{array}{cc}
        e^{i\tilde{x}_1} & 0 \\
        0 & e^{-i\tilde{x}_1}  
    \end{array}\right)\cdots \left(\begin{array}{cc}
        e^{i\tilde{x}_n} & 0 \\
        0 & e^{-i\tilde{x}_n}
    \end{array}\right)\left(\begin{array}{cc}
        1\\
        1 
    \end{array}\right)\right].
\end{align}
This expression contains complex numbers, although all entries of the tensor become real.
One can re-gauge the cores by transforming each as follows
\begin{equation}
    A_q \rightarrow T \left(T^{-1} \cdot A_q \cdot T\right) T^{-1},\quad \forall\ 1\leq q \leq n-1,
\end{equation}
with
\begin{equation}
    T = \frac{1}{\sqrt{2}}\left(\begin{array}{cc}
        1 & i \\
        i & 1 
    \end{array}\right), \quad T^{-1} =  \frac{1}{\sqrt{2}}\left(\begin{array}{cc}
        1 & -i \\
        -i & 1 
    \end{array}\right)
\end{equation}
which leads to

\begin{align}
    \cos\left(v (x-x_0)\right)  &\approx
    \frac{1}{2}\left[ \left(\begin{array}{cc}
        \frac{1+i}{\sqrt{2}}& 
        \frac{1+i}{\sqrt{2}}
    \end{array}\right)
    \left(\begin{array}{cc}
        \cos{\tilde{x}_1} & -\sin{\tilde{x}_1} \\
        \sin{\tilde{x}_1}  & \cos{\tilde{x}_1} 
    \end{array}\right)\cdots \left(\begin{array}{cc}
        \cos{\tilde{x}_n} & -\sin{\tilde{x}_n} \\
        \sin{\tilde{x}_n}  & \cos{\tilde{x}_n} 
    \end{array}\right)\left(\begin{array}{c}
       \frac{1-i}{\sqrt{2}}\\
        \frac{1-i}{\sqrt{2}}
    \end{array}\right)\right]\nonumber\\ 
    &=\frac{1}{2}\Bigg[ \left(\begin{array}{cc}
        \cos{\tilde{x}_1}+\sin{\tilde{x}_1} & 
        -\sin{\tilde{x}_1}+\cos{\tilde{x}_1} 
    \end{array}\right)\nonumber\\
    &\qquad\times
    \left(\begin{array}{cc}
        \cos{\tilde{x}_2} & -\sin{\tilde{x}_2} \\
        \sin{\tilde{x}_2}  & \cos{\tilde{x}_2} 
    \end{array}\right)\cdots
    \left(\begin{array}{cc}
        \cos{\tilde{x}_n}-\sin{\tilde{x}_n} \\
        \sin{\tilde{x}_n}+\cos{\tilde{x}_n} 
    \end{array}\right)\Bigg].
\end{align}
The $\sin$ function can be derived in the same way with a shift of $x$ by $\frac{\pi}{2}$.

\paragraph{Polynomials}
The tensor train decomposition of polynomials can be performed following \cite{Oseledets2012}.
Consider the degree-$p$ polynomial
\begin{equation}
    f(x) =\sum_{s=0}^p v_s (x-x_0)^s \approx \sum_{s=0}^p v_s \left(\sum^n_{q=1} \tilde{x}_q\right)^s,
\end{equation}
with $\tilde{x}_q=c^x_qx_q+\frac{a-x_0}{n}$.
First, we split $\tilde{x}_1$ from the rest,
\begin{align}
    f(x) &\approx \sum_{s=0}^p \sum_{t=s}^{p} \left(a_t \frac{t!}{s!(t-s)!} \tilde{x}_1^{t-s}\right)\left(\tilde{x_2}+\cdots+\tilde{x_n}\right)^s\nonumber\\
    &=C_1(x_1) \left(\begin{array}{cc}
         \left(\tilde{x}_2+\cdots+\tilde{x}_n\right)^0  \\
         \left(\tilde{x}_2+\cdots+\tilde{x}_n\right)^1\\
         \vdots\\
         \left(\tilde{x}_2+\cdots+\tilde{x}_n\right)^p
    \end{array}\right),
\end{align}
where $C_1(x_1)$ is the first core of the decomposition and is given by
\begin{equation}
    C_1(x_1)=\left(\begin{array}{ccccc}
       \sum_{t=0}^{p} a_t \tilde{x}_1^{t} & \cdots & \sum_{t=s}^{p} a_t \frac{t!}{s!(t-s)!} \tilde{x}_1^{t-s} & \cdots  &  a_p
    \end{array}\right),
\end{equation}
which is a $1\times (p+1)$ matrix, such that the rank is $p+1$.
This procedure can be iterated, leading to the final decomposition
\begin{equation}
    f = C_1(x_1)\cdot C_2(x_2)\cdots C_n(x_n)
\end{equation}
with the $(p+1)\times (p+1)$ matrices $C_q(x_q)$ for $q\geq 2$ given by
\begin{equation}
    C_q^{st}(x_q) = \begin{cases} \frac{t!}{s!(t-s)!}\tilde{x}_q^{t-s}, & \mbox{if } t \geq s \\ 0 & \mbox{otherwise,} \end{cases}
\end{equation}
and the last core
\begin{equation}
    C_n(x_n) = \left(\begin{array}{cccc}
         1  &
         \tilde{x}_n &
         \cdots&
         \tilde{x}_n^p
    \end{array}
    \right)^T.
\end{equation}

\subsection{Shift and Toeplitz matrices}
Shift matrices are defined as the operator that shift one-dimensional functions by a predefined value,
\begin{equation}
    S_d f(x) = f(x + d),
\end{equation}
where $d$ is an integer $d\in [0,N]$ specifying how much the function (or data) is shifted.
Following \cite{Kazeev2013} we define $N\times N$ shift matrices $Q^N_d$ and $R^N_d$,
\begin{equation}
    Q^N_d(i,j) = \begin{cases} 1, & \mbox{if } i - j = d \\ 0 & \mbox{otherwise,} \end{cases}
\end{equation}
and $R^N_d=\left(Q^N_{N-d}\right)^T$.
Note, that in the above definition we implicitly define $Q^N_{d}\equiv 0$ if $d>N$.
The tensor train format of $Q_d$ and $R_d$ has been derived in \cite{Kazeev2013} for the case of dimensions which are powers of two ($N=2^{n}$).
The crucial step to generalize that result is the following observation on the decomposition of shift matrices. 
Given a decomposition of $N=p\cdot q$, where $p$ and $q$ are integers, we have
\begin{align}
    Q^N_d&=Q^p_{\lfloor d/q\rfloor}\otimes Q^q_{d-q\cdot\lfloor d/q\rfloor}+Q^p_{\lfloor d/q\rfloor+1}\otimes R^q_{d-q\cdot \lfloor d/q\rfloor}\nonumber\\
    R^N_d&=R^p_{\lfloor d/q\rfloor}\otimes R^q_{d-q\cdot\lfloor d/q\rfloor}+R^p_{\lfloor d/q\rfloor+1}\otimes Q^q_{d-q\cdot \lfloor d/q\rfloor}.
\end{align}
These equations can be rewritten in terms of the rank product
\begin{equation}
    \left(\begin{array}{c}
        Q^N_d \\
         R^N_d
    \end{array}\right) = \left(\begin{array}{cc}
        Q^p_{\lfloor d/q\rfloor} &Q^p_{\lfloor d/q\rfloor+1} \\
        R^p_{\lfloor d/q\rfloor}& R^p_{\lfloor d/q\rfloor+1}
    \end{array}\right)\boxtimes \left(\begin{array}{c}
        Q^q_{d-q\cdot\lfloor d/q\rfloor} \\
         R^q_{d-q\cdot\lfloor d/q\rfloor}\end{array}\right).\label{eq:shif_dec}
\end{equation}
Using the above equation iteratively on itself, one can derive the tensor train decomposition.
We first factor $d$ into the chosen basis, $d=\sum_q c_q d_q\left(=\sum_q c^{i}_q d_q=\sum_q c^{j}_q d_q\right)$, where we assume that $i$ and $j$ have been factorized in the same way and therefore $c_q\equiv c_q^i=c_q^j$.
The digit values $d_q$ of the decomposed $d$ can be calculated recursively
\begin{equation}
    d_1 = \lfloor d/c_1 \rfloor, \quad d_i = \Bigl\lfloor (d-\prod_{j=1}^{i-1} d_j c_j)/c_i \Bigr\rfloor\ {\rm for }\ i>1. 
\end{equation}
Using Eq.~\eqref{eq:shif_dec} iteratively one reaches the final factorization:
\begin{equation}
    \left(\begin{array}{c}
        Q^N_d \\[0.3em]
         R^N_d
    \end{array}\right) = \left(\begin{array}{cc}
        Q^{b_1}_{d_1} &Q^{b_1}_{d_1+1} \\[0.3em]
        R^{b_1}_{d_1}& R^{b_1}_{d_1+1}
    \end{array}\right)\boxtimes \left(\begin{array}{cc}
        Q^{b_2}_{d_2} &Q^{b_2}_{d_2+1}  \\[0.3em]
        R^{b_2}_{d_2} & R^{b_2}_{d_2+1} 
    \end{array}\right)\boxtimes \cdots \boxtimes\left(\begin{array}{c}
        Q^{b_n}_{d_n} \\[0.3em]
        R^{b_n}_{d_n}
    \end{array}\right).
\end{equation}
The shift matrices $Q$, resp. $R$, can be obtained by performing another rank multiplication with the vector $(1,0)$, resp. $(0,1)$, from the left.
From the shift matrices one can, as described in \cite{Kazeev2013}, derive the multi-level Toeplitz matrix by stacking all shift matrices for all possible values of $d$ into one single tensor of with three sets of indices, $i=(i_1,\dotsc i_n)$, $j=(j_1,\dotsc j_n)$ and $k=(k_1,\dotsc k_n)$.
From this Toeplitz tensor one can efficiently calculate convolutions.

\subsection{Fourier transformation}
\begin{figure}[t!]
    \centering
    \includegraphics[width=0.5\linewidth]{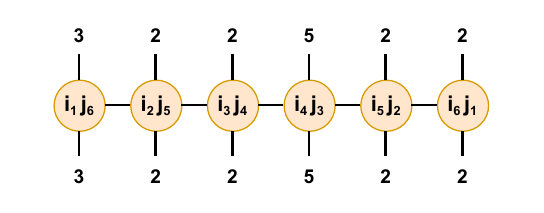}
    \caption{Shape of the Fourier tensor train in graphical notation. Here the matrix $M(i,j)$ of shape $240\times240$ is factorized into $i=80i_1+40i_2+20i_3+4i_4+2i_5+i_6$ and $j=120j_1+60j_2+12j_3+6j_4+3j_5+j_6$} ($i$ indices on the top, $j$ indices on the bottom).
    \label{fig:fourier_shape}
\end{figure}

The fact that the matrix defining the discrete Fourier transformation leads to tensor trains with small ranks has been shown in \cite{Chen2023}.
Here we generalize the concept to arbitrarily sized matrices.
We start with the definition of the square matrix of size $N$
\begin{equation}
    M(i,j) = ve^{i j},
\end{equation}
where $v=e^\frac{-2\pi i}{N}$ (here with $i$ as the imaginary number).
Using factorized indices $i=\sum^n_{q=1}c_q^ii_q$ and $j=\sum^n_{r=1}c^j_rj_r$ we get
\begin{equation}
    M(i,j) = ve^{\sum^n_{q=1}c_q^ii_q \sum^n_{r=1}c^j_rj_r}.
\end{equation}
The order of the index factorization of $i$ and $j$ is chosen in such a way that they share the same digit variables but with reversed factors associated to them.
This order has also been used in \cite{Chen2023} for power-of-two dimensions and was crucial to derive a low-rank decomposition.
Specifically, in this order each digit $j_q$ is placed on the same core as $j_{n-q+1}$ leading to a decomposition
\begin{equation}
    M(i,j) = C_1^{\mu_1}(i_1,j_n)C_2^{\mu_1\mu_2}(i_2,j_{n-1}) \dotsc  C_1^{\mu_{n-1}}(i_n,j_1).
\end{equation}
An example is depicted in Figure \ref{fig:fourier_shape}.
We continue with deriving the decomposition
\begin{equation}
    M(i,j) = v \prod^n_{q=1}\left(\prod^n_{r=1}e^{c^i_q c^j_r i_q j_r}\right)\equiv  v \prod^n_{q=1}T(i_q,j),
\end{equation}
where
\begin{equation}
    T(i_q,j) = \left(\prod^n_{r}e^{c^i_q c^j_r i_q j_r}\right) =\sum_{s=0}^{b_q^i-1} \delta(s,i_q) \prod_{r=1}^n e^{s \times c^i_q c^j_r j_r},
\end{equation}
where $\delta(s,i_q)$ is the Kronecker delta.
Note, that $T(i_q,j)$ in this form is written as a sum over $b_q^i$ terms, where each term is a rank-one tensor train, cf. Eq. \eqref{eq:exp_dec}.
The full $M(i,j)$ can therefore be written as an element-wise product over $n$ tensor trains, each with rank $b_q^i$.
This multiplication can be done with a zip-up algorithm and is only stable for the described shape where the $i$- and $j$-digits are shifted against each other.

Although our construction is different compared to the one from \cite{Chen2023}, we observe that it leads to an efficient low-rank approximation.
For instance, a matrix of size $2^{13}$ leads to distances between the exact discrete Fourier  transformation and the tensorized Fourier transformation with the values $[2.6\times10^1,3.5\times10^{-1},4.0\times10^{-6},9.3\times10^{-11}]$ for the respective maximum ranks $[2, 4, 8, 16]$ using double precision arithmetic.

\section{Solver and other algorithms} \label{sec:solver}

Besides the arithmetic, \texttt{trainsum} also implements the most common tensor structured solver for eigenvalue equations and systems of linear equations.
Eigenvalue equations are solved with the density matrix renormalization group ansatz (DMRG) \cite{White1992}, where the global eigenvalue problem collapses into local eigenvalue problems by exploiting the canonical formats \cite{Schollwoeck2011}.
The same strategy is also employed for the solution of linear equation systems.
\texttt{trainsum} implements two types of solver described in \cite{Dolgov2014}:
First, a DMRG-like approach for variationally optimizing the energy function.
Second, the so called AMEn-solver, which uses single-core sweeping strategies together with gradient enrichment.

In addition to that, \texttt{trainsum} also has an algorithm for finding the maximum and minimum values of a tensor train that was proposed in \cite{Chertkov2022}.

\section{User Guide} \label{sec:user_guide}
In this section we describe all core functionalities of \texttt{trainsum}.
The source code can be found on GitHub\footnote{\url{https://github.com/fh-igd-iet/trainsum}} while the documentation can be found at Read the Docs\footnote{\url{https://trainsum.readthedocs.io}}.

\paragraph{Installation and Import}
The package can be installed using pip via
\begin{lstlisting}[language=bash]
pip install trainsum
\end{lstlisting}
It depends on the packages \texttt{numpy}, \texttt{opt\_einsum}, \texttt{array\_api\_compat} and \texttt{h5py}.
The public API of \texttt{trainsum} is provided by the \texttt{TrainSum} class. Three backends are supported out of the box:
\begin{lstlisting}[language=python]
from trainsum.numpy import trainsum as ts
from trainsum.cupy import trainsum as ts
from trainsum.torch import trainsum as ts    
\end{lstlisting}
Since the package uses internally the Array API standard together with \texttt{opt\_einsum} and a small compatibility layer in form of the \texttt{array\_api\_compat} module one can also use other libraries.
To do so the \texttt{TrainSum} class can be initialized with an Array API conforming library.
\begin{lstlisting}[language=python]
import NDArrayLib as lib
from trainsum import TrainSum
ts = TrainSum[lib.NDArray](lib)
\end{lstlisting}
For an improved static analysis \texttt{TrainSum} is a generic class and takes the N-dimensional array class as type specifier.
\texttt{TrainSum} expects \texttt{NDArrayLib} to have mutable $N$-dimensional arrays with shapes that are defined at all times (ruling out JAX or Dask).

\paragraph{Factorized dimensions}
The most important concept of the \texttt{trainsum} module is the factorized dimension, which is expressed as the \texttt{Dimension} class.
\texttt{Dimension} is a sequence of \texttt{Digit} instances that resemble the factorization $i=\prod^{n}_{q=1}c^i_qi_q$, so that each digit can be assigned to one $i_q$.
In addition to the size of $i_m$, which is accessible by \texttt{Digit.base}, a digit is also associated with the factor $c^i_q$ \texttt{Digit.factor}.

\begin{figure}[h!]
    \centering
    \includegraphics[width=0.6\linewidth]{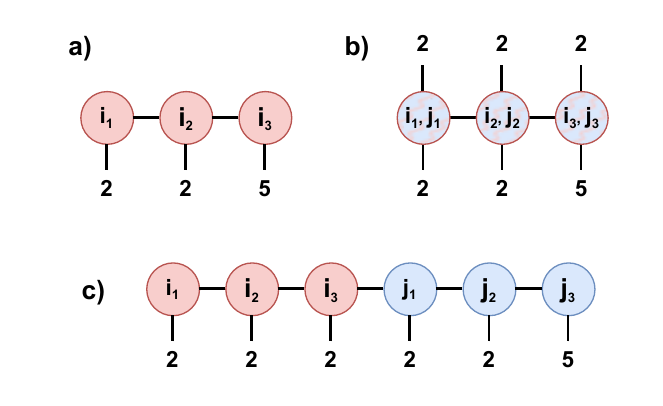}
    \caption{Exemplary tensor train shapes. a) represents a quantized vector with 20 entries. b) and c) represent a matrix of shape (8, 20). b) is the "interleaved" shape, while c) is the "block" shape.}
    \label{fig:shapes}
\end{figure}

The preferred way of constructing a \texttt{Dimension} is the \texttt{TrainSum.dimension} function.
If a integer is provided, it is factorized using a prime factor decomposition.
A sequence of integers is interpreted as the bases of the digits:
\begin{lstlisting}[language=python]
dim = ts.dimension(20) # decomposed to 2*2*5
dim = ts.dimension([2, 5, 2]) # explicit factorization 2*5*2
\end{lstlisting}
Beyond serving as a container for digits, \texttt{Dimension} is also able to compute \eqref{eq:quantization} in both directions.
Indices $i$ can be converted to digit indices $i_q$ and vice versa:
\begin{lstlisting}[language=python]
inp = np.arange(1024)
dim = ts.dimension(1024)

digit_idxs = dim.to_digits(inp)
dim_idxs = dim.to_idxs(digit_idxs) # equal to inp
\end{lstlisting}

\paragraph{Domains and uniform grids}
As shown in Section \ref{sec:structured_tensors}, one can define an interval on which discretized functions are defined.
For doing that, we provide the class \texttt{Domain}, with the \texttt{lower} and \texttt{upper} properties.
Together with a \texttt{Dimension} it is possible to define N-dimensional uniformly spaced grids:
\begin{lstlisting}[language=python]
dims = ts.dimension(1024), ts.dimension(32)
domains = ts.domain(-1.0, 1.0), ts.domain(-10.0, 10.0)
grid = ts.uniform_grid(dims, domains)
\end{lstlisting}
The class \texttt{UniformGrid} is able to convert indices of dimensions $i$ into their respective coordinates and vice versa:
\begin{lstlisting}[language=python]
dim = ts.dimension(1024)
domain = ts.domain(-1.0, 1.0)
grid = ts.uniform_grid(dim, domain)
idxs = np.arange(1024)
coords = grid.to_coords(idxs) # same as np.linspace(-1.0, 1.0, 1024)
idxs = grid.to_idxs(coords)
\end{lstlisting}

\paragraph{Shape of tensor trains}
The shape of a tensor normally describes how many dimensions the tensor has and how big the dimensions are.
It is usually represented as a tuple of integers.
Representing the shape of a tensor train is a bit more intricate.
\texttt{trainsum} provides with \texttt{TrainShape} an own class for representing the shape of a tensor train.
It has the property \texttt{dims}, which is a sequence of \texttt{Dimension} instances.
In addition to that, \texttt{TrainShape.digits} is a sequence of grouped \texttt{Digit} instances.
The \texttt{Digit} groups resemble the cores of a tensor train and are all part of its dimensions.
Beside the digits that define the open indices of a tensor train, \texttt{TrainShape} also holds the sizes of the rank dimensions.
They can be accessed with the \texttt{ranks} property or with the \texttt{rank\_left} or \texttt{rank\_right} functions.
A \texttt{TrainShape} can be created with the \texttt{TrainSum.trainshape} function.
\begin{lstlisting}[language=python]
shape = ts.trainshape(20)

dims = ts.dimension(8), 20
block_shape = ts.trainshape(*dims, mode='block')

dims = ts.dimension(8), ts.dimension(20)
interleaved_shape = ts.trainshape(*dims, mode='interleaved')
\end{lstlisting}
The different shapes are displayed in figure \ref{fig:shapes}.
The shape can also be explicitly defined:
\begin{lstlisting}[language=python]
dims = ts.dimension(8), ts.dimension(20)
block_shape = ts.trainshape(*dims, digits=[(d,) for dim in dims for d in dim])
interleaved_shape = ts.trainshape(*dims, digits=[dgts for dgts in zip(*dims)])
\end{lstlisting}

\paragraph{Arithmetic options}
A context manager in Python is a class with the magic methods \texttt{\_\_enter\_\_} and \texttt{\_\_exit\_\_}.
They are used together with the \texttt{with} statement, where \texttt{\_\_enter\_\_} is called upon reaching the \texttt{with}-line and \texttt{\_\_exit\_\_} upon the end of the \texttt{with}-scope.
Since the same arithmetic operations can be performed with multiple methods, as can be seen in Section \ref{sec:arithmetic}, \texttt{trainsum} uses context manager together with some global dictionary to define the behavior of such operations.
There are three main types of operations: Einsum operations, cross operations and evaluation operations.
Within a scope there can be only one manager per type.
The manager are stored in the dictionary with the key \texttt{(namespace, operation\_type, thread\_id)} and can be created with the \texttt{TrainSum} functions \texttt{exact}, \texttt{decomposition}, \texttt{variational}, \texttt{cross} and \texttt{evaluation}.
An exemplary usage would be:
\begin{lstlisting}[language=python]
with ts.exact():
    # ... some einsum operations that are performed exact
\end{lstlisting}

\paragraph{Tensor trains}
The main class for dealing with tensor trains in the \texttt{trainsum} package is unsurprisingly \texttt{TensorTrain}.
It combines the shape of a tensor train with the data of the cores, which can be accessed via \texttt{shape} and \texttt{cores}.
As with normal tensors, the data type and device can be accessed and altered by the \texttt{dtype} and \texttt{device} properties.
In addition to that, \texttt{TensorTrain} has some utility functions:
\begin{lstlisting}[language=python]
TensorTrain.to_tensor   # compute the full tensor
TensorTrain.extend      # extend by another tensor train
TensorTrain.conj        # return the complex conjugated
TensorTrain.normalize   # normalize upon a core
TensorTrain.truncate    # reduce the ranks using einsum options
TensorTrain.transform   # perform an element-wise operation using cross options
\end{lstlisting}
Since the corresponding magic functions like \texttt{\_\_mul\_\_} or \texttt{\_\_add\_\_} are implemented, it is also possible to do normal arithmetic with \texttt{TensorTrain}.
Addition and einsum-like operations like \texttt{\_\_mul\_\_} and \texttt{\_\_matmul\_\_} are influenced by the current einsum options.
\texttt{\_\_pow\_\_} is implemented using cross interpolation, except for the special case where a power of two is requested, which is then executed as normal multiplication.
\texttt{\_\_abs\_\_} and \texttt{\_\_truediv\_\_} are also based on cross interpolation.
Here we show an example use:
\begin{lstlisting}[language=python]
train1    = # ... 1D-tensor train
train2    = # ... 1D-tensor train
mat_train = # ... 2D-tensor train with high rank

# operations using einsum options
with ts.decomposition(max_rank=5, cutoff=1e-10):
    mul_res = train1 * train2
    add_res = train1 + train2
    train1 += train2
    pow_res = train2**2
    mat_train.truncate()
    matmul_res = mat_train @ (train1+train2)

# operations using cross options
with ts.cross(max_rank=16, eps=1e-10):
    pow_res = train1**0.5
    abs_res = abs(train2)
    div_res = train1 / train2
    func_res = div_res.transform(lambda x: np.exp(-x**2))
\end{lstlisting}
\texttt{\_\_getitem\_\_} allows the evaluation of the tensor train by the provided indices.

\paragraph{Construction}
Having provided the definitions of \texttt{Dimension}, \texttt{Domain}, \texttt{UniformGrid}, \texttt{TrainShape} and \texttt{TensorTrain} it is now time to turn towards the construction of tensor trains.
There are two different ways how such a construction can be done.
The first way is the explicit construction of functions that are supported by \texttt{trainsum}:
\begin{multicols}{2}
    \begin{itemize}
        \item constant functions
        \item exponentials
        \item cosine and sine
        \item polynomials
        \item shift matrices
        \item Toeplitz tensors
        \item transformation matrix of the discrete Fourier transform
    \end{itemize}
\end{multicols}
For the exponentials, the trigonometric functions and the polynomials it is necessary to provide a one-dimensional grid:
\begin{lstlisting}[language=python]
dim = ts.dimension(600)
domain = ts.domain(-1.0, 1.0)
grid = ts.uniform_grid(dim, domain)

coeffs = [1.0, 0.0, 0.1]
f = 1.0
x0 = 0.5

exp = ts.exp(grid, f, x0)           # exp(a(x-x0))
sin = ts.sin(grid, f, x0)           # sin(f(x-x0))
cos = ts.sin(grid, f, x0)           # cos(f(x-x0))
poly = ts.polyval(grid, coeffs, x0) # 1.0*(x-x0)**2 + 0.1
\end{lstlisting}
For the other functions a shape or a dimension is sufficient:
\begin{lstlisting}[language=python]
shape = ts.trainshape(500)
dim = shape.dims[0]

full = ts.full(shape, 1.0)                  # f(x)=v
shift = ts.shift(dim, 1, circular=False)    # eye shifted by one
toep = ts.toeplitz(dim, mode='full')        # toeplitz tensors
qft = ts.qft(dim)                           # fourier matrix
\end{lstlisting}

A more flexible construction can be achieved by using the \texttt{TrainSum.tensortrain} function.
The first argument is always a \texttt{TrainShape}.
The second argument can be either a tensor, a function or a sequence of tensors.
If it is a tensor, the tensor is approximated by the currently active einsum options.
A function is approximated by a cross interpolation using the cross options, while a sequence of tensors is interpreted as the cores of the tensor train.
\begin{lstlisting}[language=python]
shape = ts.trainshape(720)
domain = ts.domain(-1.0, 1.0),
grid = ts.uniform_grid(shape.dims, domain)

# data approximation
data = np.linspace(-1.0, 1.0, shape.dims[0].size())**2
with tt.variational(max_rank=8, cutoff=0.0, ncores=2, nsweeps=2):
    train = ts.tensortrain(shape, data)

# function approximation
func = lambda idxs: np.exp(-np.sum(grid.to_coords(idxs)**2, axis=0))
with tt.cross(max_rank=8, eps=1e-10):
    train = ts.tensortrain(shape, func)

# explicit construction with cores
cores = [np.ones([1, digit.base, 1]) for digit in shape.dims[0]]
train = ts.tensortrain(shape, cores)
\end{lstlisting}

\paragraph{Arithmetic}
Beyond the arithmetic of the \texttt{TensorTrain} class there exist a few functions in \texttt{TrainSum} that are noteworthy.
To start with, there is the \texttt{einsum} function which offers the same functionality as \texttt{numpy.einsum}.
\texttt{einsum} takes a string defining the Einstein summation together with the corresponding \texttt{TensorTrains}.
In addition to \texttt{einsum}, there is the function \texttt{einsum\_expression}, which takes \texttt{TrainShapes} instead of \texttt{TensorTrains} and returns a \texttt{Callable}.
The \texttt{Callable} can be again called with \texttt{TensorTrains} but without the overhead of optimizing all the internal Einstein summations.
The \texttt{Callable} uses the einsum options from where it is initialized not from where it is called.
\begin{lstlisting}[language=python]
shape = ts.trainshape(1024)
train = ts.full(shape, 1.0)
mat = ts.shift(shape.dims[0], 512)

with ts.exact():
    res = ts.einsum('ij,j->i', mat, train)
    expr = ts.einsum_expression('ij,j->i', mat.shape, train.shape)
res = expr(mat, train) # using the exact options
\end{lstlisting}

\texttt{evaluate} and \texttt{evaluate\_expression} offer the same functionalities as \texttt{einsum} and \texttt{einsum\_expression} with the difference, that there is an additional argument as input, which is an index tensor that should be evaluated.
The functions are influenced by the evaluation options and the evaluation is performed exactly.

\texttt{add} is another interesting function, as it offers the addition of multiple tensor trains at once.
In the end, there is \texttt{min\_max}, a function that is able to approximately calculate the minimum and maximum values and indices of a tensor train.
In addition to the tensor train, \texttt{min\_max} takes an integer as argument defining how many indices should be temporarily used.
The more indices are used the more accurate the result will be.

\paragraph{Solver}
There are two kinds of solvers in \texttt{trainsum}: Solver for eigenvalue equations and solver for linear equation systems.
Both solver depend on the \texttt{LinearMap} class, that resembles a linear map for some multidimensional tensor train.
As a first argument it takes a string defining the linear map as an Einstein summation.
Consecutive arguments are the tensor trains of the operation.
The tensor train, that the linear operator acts upon, is identified by a \texttt{TrainShape}.
\begin{lstlisting}[language=python]
op1   = # ... 2D tensor train
op2   = # ... 2D tensor train
shape = # ... 1D tensor train shape

lin_map = ts.LinearMap("ij,jl,l->i", op1, op2, shape)
\end{lstlisting}
The requirements of the linear map are, that the result has the same shape as the provided input shape and that all digits of the input and output dimensions reside on the same core.

With the definition of \texttt{LinearMap} it is possible to define a solver for an eigenvalue equation $H\Psi=E\Psi$.
\begin{lstlisting}[language=python]
lin_map = # ... some LinearMap
guess   = # ... a guess state

# define the options
local_solver = ts.lanczos()
decomposition = ts.svdecomposition(max_rank=10)
strategy = ts.sweeping_strategy(ncores=2, nsweeps=10)

# construct the solver
solver = ts.eigsolver(
    lin_map,
    decomposition=decomposition,
    strategy=strategy,
    solver=local_solver)

# call the solver for starting the solving process
result = solver(guess)
\end{lstlisting}
As can be seen the solver depends on a local eigenvalue solver, a sweeping strategy and for strategies with \texttt{ncores>=2} on a matrix decomposition to deal with occuring super-cores.
Linear solver are equally defined and only require an additional tensor train for the right hand side of the equation $Ax=b$ and a local linear equation solver provided by the function $\texttt{gmres}$.

\paragraph{Examples}
We provide detailed examples for use cases in the fields of simulation and data analysis including:
\begin{itemize}
    \item Solving a system of linear equations from a PDE (heat equation) or eigenvalue equations (hydrogen atom) in the finite difference formulation,
    \item Fourier transformations for spectra,
    \item Picture compression,
    \item Tensor trains in neural networks for picture classification (MNIST data),
    \item Discrete convolutions using Toeplitz tensors.
\end{itemize}
The examples are provided as Jupyter notebooks and are part of the documentation at Read the Docs.\footnote{\url{https://trainsum.readthedocs.io}}

\section{Conclusion}  \label{sec:conclusion}
We presented \texttt{trainsum}, an easy-to-use software package written in Python for dealing with quantics tensor trains.
The two main features making it a valuable addition to existing software are on the one hand, that dimensions of arbitrary sizes can be factroized using prime factorization and on the other hand, that Einstein summations can be easily performed for multidimensional tensor trains.
In addition to that \texttt{trainsum} is mostly based on Python's Array API standard and \texttt{opt\_einsum}, which allows it to be used with different backends including NumPy, Torch and CuPy.
Since there are different methods for performing arithmetic operations when dealing with tensor networks, \texttt{trainsum} makes heavy use of context managers to provide an easy and variable way for defining tensor train arithmetic.

Although only a small step into this direction, the generalizations done in this paper might prove fruitful for creating a tensor network library that resembles the same functionalities as other N-dimensional array libraries like NumPy.
One of the missing parts for creating such a library are arbitrary slicing operators, which could be used for assignment operations.
In addition to that, the stability of some of the implemented algorithms, especially the cross interpolation, could be improved, which is a topic of ongoing research.

\section*{Acknowledgements}
This work was supported by the research project "Zentrum für Angewandtes Quantencomputing" (ZAQC), which is funded by the Hessian Ministry for Digital Strategy and Innovation and the Hessian Ministry of Higher Education, Research and the Arts.

\clearpage

\bibliographystyle{ieeetr}
\bibliography{references}

@Article{NunezFernandez2025,
  author    = {Núñez Fernández, Yuriel and Ritter, Marc K and Jeannin, Matthieu and Li, Jheng-Wei and Kloss, Thomas and Louvet, Thibaud and Terasaki, Satoshi and Parcollet, Olivier and von Delft, Jan and Shinaoka, Hiroshi and Waintal, Xavier},
  journal   = {SciPost Physics},
  title     = {Learning tensor networks with tensor cross interpolation: New algorithms and libraries},
  year      = {2025},
  issn      = {2542-4653},
  month     = mar,
  number    = {3},
  volume    = {18},
  doi       = {10.21468/scipostphys.18.3.104},
  publisher = {Stichting SciPost},
}

@Article{Schollwoeck2011,
  author    = {Schollwöck, Ulrich},
  journal   = {Philosophical Transactions of the Royal Society A: Mathematical, Physical and Engineering Sciences},
  title     = {The density-matrix renormalization group: a short introduction},
  year      = {2011},
  issn      = {1471-2962},
  month     = jul,
  number    = {1946},
  pages     = {2643--2661},
  volume    = {369},
  doi       = {10.1098/rsta.2010.0382},
  publisher = {The Royal Society},
}

@Article{Dolgov2014,
  author    = {Dolgov, Sergey V. and Savostyanov, Dmitry V.},
  journal   = {SIAM Journal on Scientific Computing},
  title     = {Alternating Minimal Energy Methods for Linear Systems in Higher Dimensions},
  year      = {2014},
  issn      = {1095-7197},
  month     = jan,
  number    = {5},
  pages     = {A2248--A2271},
  volume    = {36},
  doi       = {10.1137/140953289},
  publisher = {Society for Industrial & Applied Mathematics (SIAM)},
}

@Misc{Chertkov2022,
  author    = {Chertkov, Andrei and Ryzhakov, Gleb and Novikov, Georgii and Oseledets, Ivan},
  title     = {Optimization of Functions Given in the Tensor Train Format},
  year      = {2022},
  copyright = {arXiv.org perpetual, non-exclusive license},
  doi       = {10.48550/ARXIV.2209.14808},
  publisher = {arXiv},
}

@Article{Stoudenmire2010,
  author    = {Stoudenmire, E M and White, Steven R},
  journal   = {New Journal of Physics},
  title     = {Minimally entangled typical thermal state algorithms},
  year      = {2010},
  issn      = {1367-2630},
  month     = may,
  number    = {5},
  pages     = {055026},
  volume    = {12},
  doi       = {10.1088/1367-2630/12/5/055026},
  publisher = {IOP Publishing},
}

@Article{Paeckel2019,
  author    = {Paeckel, Sebastian and Köhler, Thomas and Swoboda, Andreas and Manmana, Salvatore R. and Schollwöck, Ulrich and Hubig, Claudius},
  journal   = {Annals of Physics},
  title     = {Time-evolution methods for matrix-product states},
  year      = {2019},
  issn      = {0003-4916},
  month     = dec,
  pages     = {167998},
  volume    = {411},
  doi       = {10.1016/j.aop.2019.167998},
  publisher = {Elsevier BV},
}

@Article{Schollwoeck2011a,
  author    = {Schollwöck, Ulrich},
  journal   = {Annals of Physics},
  title     = {The density-matrix renormalization group in the age of matrix product states},
  year      = {2011},
  issn      = {0003-4916},
  month     = jan,
  number    = {1},
  pages     = {96--192},
  volume    = {326},
  doi       = {10.1016/j.aop.2010.09.012},
  publisher = {Elsevier BV},
}

@Article{Banuls2023,
  author    = {Bañuls, Mari Carmen},
  journal   = {Annual Review of Condensed Matter Physics},
  title     = {Tensor Network Algorithms: A Route Map},
  year      = {2023},
  issn      = {1947-5462},
  month     = mar,
  number    = {1},
  pages     = {173--191},
  volume    = {14},
  doi       = {10.1146/annurev-conmatphys-040721-022705},
  publisher = {Annual Reviews},
}

@Misc{Wang2023,
  author    = {Wang, Maolin and Pan, Yu and Xu, Zenglin and Li, Guangxi and Yang, Xiangli and Mandic, Danilo and Cichocki, Andrzej},
  title     = {Tensor Networks Meet Neural Networks: A Survey and Future Perspectives},
  year      = {2023},
  copyright = {arXiv.org perpetual, non-exclusive license},
  doi       = {10.48550/ARXIV.2302.09019},
  publisher = {arXiv},
}

@Article{Berezutskii2025,
  author    = {Berezutskii, Aleksandr and Liu, Minzhao and Acharya, Atithi and Ellerbrock, Roman and Gray, Johnnie and Haghshenas, Reza and He, Zichang and Khan, Abid and Kuzmin, Viacheslav and Lyakh, Dmitry and Lykov, Danylo and Mandrà, Salvatore and Mansell, Christopher and Melnikov, Alexey and Melnikov, Artem and Mironov, Vladimir and Morozov, Dmitry and Neukart, Florian and Nocera, Alberto and Perlin, Michael A. and Perelshtein, Michael and Steinberg, Matthew and Shaydulin, Ruslan and Villalonga, Benjamin and Pflitsch, Markus and Pistoia, Marco and Vinokur, Valerii and Alexeev, Yuri},
  journal   = {Nature Reviews Physics},
  title     = {Tensor networks for quantum computing},
  year      = {2025},
  issn      = {2522-5820},
  month     = jul,
  number    = {10},
  pages     = {581--593},
  volume    = {7},
  doi       = {10.1038/s42254-025-00853-1},
  publisher = {Springer Science and Business Media LLC},
}

@Article{Vidal2007,
  author    = {Vidal, G.},
  journal   = {Physical Review Letters},
  title     = {Entanglement Renormalization},
  year      = {2007},
  issn      = {1079-7114},
  month     = nov,
  number    = {22},
  pages     = {220405},
  volume    = {99},
  doi       = {10.1103/physrevlett.99.220405},
  publisher = {American Physical Society (APS)},
}

@Article{Dolgov2013,
  author    = {Dolgov, Sergey and Khoromskij, Boris},
  journal   = {SIAM Journal on Matrix Analysis and Applications},
  title     = {Two-Level QTT-Tucker Format for Optimized Tensor Calculus},
  year      = {2013},
  issn      = {1095-7162},
  month     = jan,
  number    = {2},
  pages     = {593--623},
  volume    = {34},
  doi       = {10.1137/120882597},
  publisher = {Society for Industrial & Applied Mathematics (SIAM)},
}

@Article{Kressner2014,
  author    = {Kressner, Daniel and Tobler, Christine},
  journal   = {ACM Transactions on Mathematical Software},
  title     = {Algorithm 941: htucker---A Matlab Toolbox for Tensors in Hierarchical Tucker Format},
  year      = {2014},
  issn      = {1557-7295},
  month     = apr,
  number    = {3},
  pages     = {1--22},
  volume    = {40},
  doi       = {10.1145/2538688},
  publisher = {Association for Computing Machinery (ACM)},
}

@Article{Oseledets2011,
  author    = {Oseledets, I. V.},
  journal   = {SIAM Journal on Scientific Computing},
  title     = {Tensor-Train Decomposition},
  year      = {2011},
  issn      = {1095-7197},
  month     = jan,
  number    = {5},
  pages     = {2295--2317},
  volume    = {33},
  doi       = {10.1137/090752286},
  publisher = {Society for Industrial & Applied Mathematics (SIAM)},
}

@Article{White1992,
  author    = {White, Steven R.},
  journal   = {Physical Review Letters},
  title     = {Density matrix formulation for quantum renormalization groups},
  year      = {1992},
  issn      = {0031-9007},
  month     = nov,
  number    = {19},
  pages     = {2863--2866},
  volume    = {69},
  doi       = {10.1103/physrevlett.69.2863},
  publisher = {American Physical Society (APS)},
}

@Article{Khoromskij2011,
  author    = {Khoromskij, Boris N.},
  journal   = {Constructive Approximation},
  title     = {O(dlog N)-Quantics Approximation of N-d Tensors in High-Dimensional Numerical Modeling},
  year      = {2011},
  issn      = {1432-0940},
  month     = apr,
  number    = {2},
  pages     = {257--280},
  volume    = {34},
  doi       = {10.1007/s00365-011-9131-1},
  publisher = {Springer Science and Business Media LLC},
}

@misc{ArrayStandrd,
    organization={Consortium for Python Data API Standards},
    title = {Python array API standard 2024.12},
    howpublished = {\url{https://data-apis.org/array-api/2024.12}},
    year = {2024},
    note = {Accessed: 2026-16-02}
}

@misc{torchtt,
howpublished = {\url{https://pypi.org/project/torchTT}},
year = {2024},
note = {Accessed: 2026-16-02}
}

@misc{TT_box,
author = {Oseledets, I and Dolgov, S. and Boyko, A. and Savostyanov, D. and Novikov, A. and Mach, T.},
title = {Matlab implementation of tensor train decomposition},
howpublished = {\url{https://github.com/oseledets/TT-Toolbox}},
year = {2011},
}

@Article{Fishman2022,
  author    = {Fishman, Matthew and White, Steven and Stoudenmire, Edwin},
  journal   = {SciPost Physics Codebases},
  title     = {The ITensor Software Library for Tensor Network Calculations},
  year      = {2022},
  month     = aug,
  doi       = {10.21468/scipostphyscodeb.4},
  publisher = {Stichting SciPost},
}

@article{gray2018quimb,
    title={quimb: a python library for quantum information and many-body calculations},
    author={Gray, Johnnie},
    journal={Journal of Open Source Software},
    year = {2018},
    volume={3}, number={29}, pages={819},
    doi={10.21105/joss.00819},
}

@Article{Oseledets2012,
  author    = {Oseledets, I. V.},
  journal   = {Constructive Approximation},
  title     = {Constructive Representation of Functions in Low-Rank Tensor Formats},
  year      = {2012},
  issn      = {1432-0940},
  month     = dec,
  number    = {1},
  pages     = {1--18},
  volume    = {37},
  doi       = {10.1007/s00365-012-9175-x},
  publisher = {Springer Science and Business Media LLC},
}

@Article{Kazeev2013,
  author    = {Kazeev, Vladimir A. and Khoromskij, Boris N. and Tyrtyshnikov, Eugene E.},
  journal   = {SIAM Journal on Scientific Computing},
  title     = {Multilevel Toeplitz Matrices Generated by Tensor-Structured Vectors and Convolution with Logarithmic Complexity},
  year      = {2013},
  issn      = {1095-7197},
  month     = jan,
  number    = {3},
  pages     = {A1511--A1536},
  volume    = {35},
  doi       = {10.1137/110844830},
  publisher = {Society for Industrial & Applied Mathematics (SIAM)},
}

@Article{Chen2023,
  author    = {Chen, Jielun and Stoudenmire, E.M. and White, Steven R.},
  journal   = {PRX Quantum},
  title     = {Quantum Fourier Transform Has Small Entanglement},
  year      = {2023},
  issn      = {2691-3399},
  month     = oct,
  number    = {4},
  pages     = {040318},
  volume    = {4},
  doi       = {10.1103/prxquantum.4.040318},
  publisher = {American Physical Society (APS)},
}

@Article{G.A.Smith2018,
  author    = {G. A. Smith, Daniel and Gray, Johnnie},
  journal   = {Journal of Open Source Software},
  title     = {opt\_einsum - A Python package for optimizing contraction order for einsum-like expressions},
  year      = {2018},
  issn      = {2475-9066},
  month     = jun,
  number    = {26},
  pages     = {753},
  volume    = {3},
  doi       = {10.21105/joss.00753},
  publisher = {The Open Journal},
}

@article{tntorch,
  author  = {Mikhail Usvyatsov and Rafael Ballester-Ripoll and Konrad Schindler},
  title   = {tntorch: Tensor Network Learning with {PyTorch}},
  journal = {Journal of Machine Learning Research},
  year    = {2022},
  volume  = {23},
  number  = {208},
  pages   = {1--6},
  url     = {http://jmlr.org/papers/v23/21-1197.html}
}

@article{OSELEDETS201070,
title = {TT-cross approximation for multidimensional arrays},
journal = {Linear Algebra and its Applications},
volume = {432},
number = {1},
pages = {70-88},
year = {2010},
issn = {0024-3795},
doi = {https://doi.org/10.1016/j.laa.2009.07.024},
url = {https://www.sciencedirect.com/science/article/pii/S0024379509003747},
author = {Ivan Oseledets and Eugene Tyrtyshnikov}
}

\end{document}